\documentclass[aps, prl, amsmath, amssymb]{revtex4}

\begin{document}

\title{The Mandelstam-Terning Line Integral in Unparticle Physics\\A 
Reply to Galloway, Martin and Stancato}
\author{A. Lewis Licht}
\affiliation{Dept. of Physics\\U. of Illinois at Chicago\\Chicago, 
Illinois 60607\\licht@uic.edu}

\begin{abstract}
We show that the path ordered Wilson line integral used to make a 
nonlocal action gauge invariant is mathematically inconsistent. We 
also show that it can lead to reasonable gauge field vertexes by the 
use of a second mathematically unjustifiable procedure.
\end{abstract}

\maketitle

\section{Introduction}\label{S:intro}

There have been attempts to make the non local unparticle action 
introduced by Georgi~\cite{Geo-1}~\cite{Geo-2} gauge 
invariant~\cite{Tern-1}~\cite{Tern-2} by introducing in the action 
the path ordered Wilson line
\begin{equation}\label{E:Wilson}
W_P \left( {y,x} \right) = P\exp \left[ { - ig\int_x^y {A_\mu  \left( w \right)dw^\mu  } } \right]
\end{equation}
where here
\begin{equation}
A_\mu  \left( w \right) = T^a A_\mu ^a \left( w \right)
\end{equation}
and $W_P $ is assumed to satisfy the condition introduced by 
Mandelstam~\cite{Mand}

\begin{equation}\label{E:mandle}
\frac{\partial }
{{\partial y^\mu  }}W_P \left( {y,x} \right) =  - igA_\mu  \left( y \right)W_P \left( {y,x} \right)
\end{equation}
We pointed out in~\cite{all-1} that $W_{P}$ is not well defined if the 
path between x and y is not specified. We took as the path a straight 
line and derived some rather complicated gauge-particle vertexes. 
Galloway, Martin and Stancato [GMS]~\cite{gms} have shown that 
$W_{{P}}$ with condition~(\ref{E:mandle}) does give vertexes that 
satisfy the ``Minimal Coupling'' requirement, that is, when the 
unparticle dimension is an integer, the coupling is what one would 
expect from simply replacing in the action $i\partial _\mu  $ by $i\partial _\mu   - gA_\mu  $
. They also point out that for integer dimension other than 1, the 
straight line method of Ref.~\cite{all-1} gives results that do not 
satisfy the Minimal Coupling requirement.  In a later 
work,~\cite{all-2} we showed that $i\partial _\mu   - gA_\mu  $
can be defined as a differential-integral operator, can be used in 
unparticle actions of arbitrary dimension and gives vertexes that 
automatically satisfy the Minimal Coupliing requirement.

Here we will show that the Wilson line with 
condition~(\ref{E:mandle}) is mathematically very shaky. We will also 
show why nevertheless it does seem to give reasonable vertexes.

\section{The Mandelstam Condition Theorem}\label{S:MCT}
Theorem:  If the Wilson line operator $W_{P}$ satisfies 
condition~(\ref{E:mandle}) then the field strength $F_{\mu\nu}$ 
vanishes and the gauge field $A_{\mu}$ is gauge equivalent to zero.

Proof:  The second partial derivative of $W_{P}$ gives, using 
condition~(\ref{E:mandle}),
\begin{equation}
\frac{\partial }
{{\partial y^\nu  }}\frac{\partial }
{{\partial y^\mu  }}W_P \left( {y,x} \right) =  - ig\frac{\partial }
{{\partial y^\nu  }}A_\mu  \left( y \right)W_P \left( {y,x} \right) - g^2 A_\mu  \left( y \right)A_\nu  \left( y \right)W_P \left( {y,x} \right)
\end{equation}
But, second derivatives commute, so
\begin{equation}
\begin{gathered}
  0 = \left( {\frac{\partial }
{{\partial y^\mu  }}\frac{\partial }
{{\partial y^\nu  }} - \frac{\partial }
{{\partial y^\nu  }}\frac{\partial }
{{\partial y^\mu  }}} \right)W_P \left( {y,x} \right) \\ 
   =  - igF_{\mu \nu } \left( y \right)W_P \left( {y,x} \right) \\ 
\end{gathered} 
\end{equation}
where
\begin{equation}
F_{\mu \nu }  = \partial _\mu  A_\nu   - \partial _\nu  A_\mu   + ig\left[ {A_\mu  ,A_\nu  } \right]
\end{equation}
is the field strength.  However, $W_{P}$ is unitary, therefore
\begin{equation}
F_{\mu \nu }  = 0
\end{equation}
This is the condition that $A_{\mu}$ be gauge equivalent to zero.  It 
is possible to find the transformation that takes $A_{\mu}$ into 
zero.  Eq.~(\ref{E:mandle}) can also be written as 
\begin{equation}\label{E:leftmandle}
\frac{\partial }
{{\partial x^\mu  }}W_P \left( {y,x} \right) =  + igW_P \left( {y,x} \right)A_\mu  \left( x \right)
\end{equation}
Under a gauge transformation V,
\begin{equation}\label{E:gauge}
\begin{gathered}
  \Phi _u  \to \Phi _u ' = V\Phi _u  \\ 
  A_\mu   \to A'_\mu   = VA_\mu  V^\dag   + \frac{i}
{g}\left( {\partial _\mu  V} \right)V^\dag   \\ 
  F_{\mu \nu }  \to F'_{\mu \nu }  = VF_{\mu \nu } V^\dag   \\ 
\end{gathered} 
\end{equation}
We have
\begin{equation}
\left( {i\partial _\mu   - gA_\mu  } \right)\Phi _u  \to \left( {i\partial _\mu   - gA'_\mu  } \right)\Phi '_u  = V\left( {i\partial _\mu   - gA_\mu  } \right)\Phi _u 
\end{equation}
If we take for the gauge transformation the unitary operator
\begin{equation}
V(x) = W_P \left( {0,x} \right)
\end{equation}
then Eq.~(\ref{E:leftmandle}) leads to
\begin{equation}
\partial _\mu  V = igVA_\mu  
\end{equation}
which makes the $A'_{\mu}$ of Eq.~(\ref{E:gauge}) equal to zero.  qed.

\section{Vertexes}\label{S:vertexes}

The question now arises, if the gauge field is actually zero, why 
does the Terning method lead to reasonable vertexes?  Basically this 
is a case of two wrongs making a right.  Introducing the Wilson line 
of Eq.~(\ref{E:Wilson}) into the unparticle action leads to an action 
that can be expanded in a power series in the $A_{\mu}$ field:
\begin{equation}
I = \sum\limits_{n = 0}^\infty  {\int {d^4 xd^4 y\Phi _u^\dag  \left( y \right)\prod\limits_{k = 1}^n {\int {d^4 z_k A_{\mu _k } \left( {z_k } \right)} } } } K_k \left( {x,y,\left\{ {z_i } \right\}} \right)\Phi _u \left( x \right)
\end{equation}
The vertexes are found by taking functional derivatives with respect 
to the $A_{\mu}$ field, using
\begin{equation}\label{E:deriv}
\frac{{\delta A_\alpha ^a \left( z \right)}}
{{\delta A_\beta ^b \left( w \right)}} = \delta ^{ab} \delta _\alpha ^\beta  \delta ^4 \left( {z - w} \right)
\end{equation}
However, as a pure gauge field $A_{\alpha}^{a}$ must be written in 
terms of a fleld $\Lambda^{a}$ where, with
\begin{equation}
\Lambda  = \Lambda ^a T^a 
\end{equation}
and
\begin{equation}
V = e^{ - ig\Lambda } 
\end{equation}
we get
\begin{equation}
\begin{gathered}
  A_\alpha   = \frac{i}
{g}\left( {\partial _\alpha  V} \right)V^\dag   \\ 
   = \int_0^1 {e^{ - itg\Lambda } \left( {\partial _\alpha  \Lambda } \right)e^{itg\Lambda } dt}  \\ 
\end{gathered} 
\end{equation}
For the simple case of U(1), or for more general gauge groups with infinitesimal $\Lambda$,
\begin{equation}
A_\alpha   = \partial _\alpha  \Lambda 
\end{equation}
and although we can write
\begin{equation}
\frac{{\delta \partial _\alpha  \Lambda ^a \left( z \right)}}
{{\delta \Lambda ^b \left( x \right)}} = \delta ^{ab} \frac{\partial }
{{\partial z^\alpha  }}\delta ^4 \left( {x - z} \right)
\end{equation}
there is no way to convert the $\partial /\partial z^\alpha  $
into a $\delta _\alpha ^\beta  $ and the functional derivative of 
Eq.~(\ref{E:deriv} cannot be carried out.

What is actually being done, is to replace, by hand, the actual $\partial _\mu  \Lambda ^a $
dependent terms provided by the Wilson line factor with non pure gauge $A_\mu ^a $
factors and then to differentiate.  This should give the correct 
vertexes, if, as one expects, the Wilson line factor puts the pure 
gauge factors exactly where the principal of minimal coupling says 
they should be.  This is certainly true for integer unparticle 
dimension, as GMS have shown.  

\section{Conclusions}\label{S:conclusion}
We have shown that the Mandelstam assumption about the Wilson line 
operator implies that the gauge fields are gauge equivalent to zero.  
We show that nevertheless the resulting unparticle action can be 
converted, by hand, to a form that gives the correct vertexes.

We would like to point out however that the mechanism that gives 
rise to unparticles is not precisely known.  It is conceivable that 
the unparticle effective action might not actually obey the minimal 
coupling principle.

\section{Acknowledgements}\label{S:Acknow}
I would like to thank Wai-Yee Keung for intereting me in this 
subject.  I would also like to thank Galloway, Martin and Stancato 
for their very interesting comments.

\end{document}